latex file
\documentstyle[12pt]{article}
\font\tenrm=cmr10

\begin{document}
\renewenvironment{thebibliography}[1]
  { \begin{list}{\arabic{enumi}.}
    {\usecounter{enumi} \setlength{\parsep}{0pt}
     \setlength{\itemsep}{3pt} \settowidth{\labelwidth}{#1.}
     \sloppy
    }}{\end{list}}

\parindent=1.5pc

\begin{center}{{\bf VACUUM STRUCTURE OF PURE GAUGE THEORIES\\
               \vglue 3pt
               ON THE LATTICE\footnote{Talk at 1992 Paris Conference on the
QCD Vacuum presented by R. Haymaker}}\\
\vglue 1.0cm
{Richard W. Haymaker, Vandana Singh and Dana Browne}\\
\baselineskip=14pt
{\it Department of Physics and Astronomy}\\
\baselineskip=14pt
{\it Louisiana State University, Baton Rouge, Louisiana, 70808, USA}\\
\vglue 0.3cm
{and}\\
\vglue 0.3cm
{Jacek Wosiek}\\
\baselineskip=14pt
{\it Chair of Computer Science}\\
\baselineskip=14pt
{\it Jagellonian University, Institute of Physics, Reymonta 4 Cracow, Poland;
and}\\
\baselineskip=14pt
{\it Max-Planck-Institut f\"{u}r Physik - Werner-Heisenberg-Institute -
P.O. Box 40 12 12, Munich, Germany}\\
\vglue 0.8cm
{ABSTRACT}}
\end{center}
\vglue 0.3cm
{\rightskip=3pc
 \leftskip=3pc
 \tenrm\baselineskip=12pt
 \noindent
We present results from simulations on two aspects of quark confinement
in the pure gauge sector. First is the calculation of the
profile of the flux tube connecting
a static  $q \bar{q}$ pair in $SU(2)$.  By using the Michael sum rules
as a constraint we give evidence that the energy density at the center
of the flux tube goes to a constant as a function of quark separation.
Slow variation of the width and energy density is not ruled out.
Secondly in the  confined phase of lattice $U(1)$ we calculate the
curl of the magnetic monopole current and show that
the dual London equation is satisfied
and that the electric fluxoid is quantized.
\vglue 0.8cm}
%
{\bf\noindent 1. Introduction}
\vglue 0.4cm
\baselineskip=14pt

I would like to report on efforts of the Cracow-LSU collaboration to
study the response of gauge fields to the presence of static sources.
Confinement is the central issue which is seen as a consequence of
the vacuum squeezing the field lines to form flux tubes connecting
color charges.  One is beginning to see considerable detail of the field
distributions in the flux tube, e.g. its size and shape, the chromoelectric
and chromomagnetic field components, and monopole currents that are
responsible for squeezing the tube to form an Abrikosov vortex.

In the first part of this talk I will describe our results for
SU(2) lattice gauge theory\cite{lsukrak,hw,hsw}.
We examine the flux tube
between a $q \bar{q}$ pair, calculating the six chromoelectric
and chromomagnetic components to the energy density and action density.
We parametrize the profile of the flux tube and study scaling.  The
results are subjected to the check provided by the Michael sum
rules\cite{michael1}.

The second part of this talk concerns the mechanism that
leads to flux tube formation.  This is the role of the solenoidal
magnetic monopole currents that surround the flux tube. In this
work\cite{shb} we show in the
confined phase of U(1) that the curl of the monopole current has a
profile similar to the electric field and that the dual London equation
is satisfied  and
electric fluxoid quantization occurs.  We demonstrate that the
the flux tube is precisely the dual of the Abrikosov vortex
in Type II superconducting materials.

%
\vglue 0.6cm
{\bf\noindent 2. Flux Tubes in SU(2) \footnote{Haymaker, Singh and Wosiek}}
%
\vglue 0.2cm
{\it\noindent 2.1. Background}
\vglue 0.1cm
\baselineskip=14pt

In this calculation we measure the field energy densities by
correlating the small plaquette with the Wilson loop.
Full details of the simulation are given in Ref. \cite{hw},
which also contains further references.  Further details of the
flux profiles will appear in a companion paper Ref. \cite{hsw}.
By fixing our attention on the middle time slice of the Wilson loop,
the time-like segments form world lines that approximate a static
$q \bar{q}$ pair. The 3 space-space plaquettes measure the magnetic
component of the energy density and similarly the 3 space-time plaquettes
measure the electric components.

Before defining the flux calculation in more detail, we point
out that the Wilson loops themselves are used to
extract the transfer matrix eigenvalues which give the static quark
potential and are further used to extrapolate the flux measurements
to infinite time extent of the Wilson loops.    Specifically we
determine the eigenvalues of the transfer matrix by fitting the
Wilson loops to the exponentials as described in Ref.\cite{hw}.
\begin{eqnarray}
<W(R,T)> = \sum_i A_i e^{-E_i(R) T}.
\label{e10}
\end{eqnarray}
$E_0(R)$ is of special interest since it contains the static
quark potential:
\begin{eqnarray}
E_0(R) = -\frac{\alpha}{R} + \sigma R + \frac{c(\beta)}{a(\beta)}.
\label{e20}
\end{eqnarray}
The term independent of $R$ is the self energy of the two quarks which
does not scale but diverges as $a \rightarrow 0$.  Since Wilson loops
can be calculated quite accurately, the static potential is a useful
physical quantity to check scaling and thereby determine the lattice
spacing $a(\beta)$.  All our data is consistent with standard values
$a(2.3) = 0.171$ fm, $a(2.4) = 0.128$ fm and $a(2.5) = 0.089$ fm.
\vglue 0.4cm
%
{\it \noindent 2.2. Flux Tube Profiles}
\vglue 0.1cm
The lattice observable needed to measure the flux is
the following\cite{back,lsukrak,hw}.
\begin{eqnarray}
    f^{\mu \nu}(x) &=& \frac{\beta}{a^4} \left(
\frac{\langle W P^{\mu \nu}_x \rangle }{\langle W \rangle}
-  \langle P \rangle \right),
\nonumber\\
 &\approx&  \frac{\beta}{a^4} \left(
\frac{\langle W P^{\mu \nu}_x - W P^{\mu \nu}_{x_{R}} \rangle}
{\langle W \rangle}  \right),
\label{e30}
\end{eqnarray}
where $W$ is the Wilson loop, $P^{\mu \nu}_x$ the plaquette
located at $x$, $\beta = \frac{4}{g^2}$ and $x_{R}$ is a distant
reference point.  In the classical continuum limit
\begin{equation}
  f^{\mu \nu} \stackrel{a \rightarrow 0} {\longrightarrow} -\frac{1}
 {2} \langle ( F^{\mu \nu} )^{2} \rangle_{q\overline{q} - vac},
\label{e40}
\end{equation}
where the notation $ \langle \cdots \rangle_{q\overline{q}-vac} $ means
the difference
of the average values in the $q\overline{q}$ and vacuum state. From now
on we shall be using field components in Minkowski space and hence
\begin{equation}
  f^{\mu \nu} \rightarrow
   \frac{1}{2} ( -B_1^2,-B_2^2,-B_3^2; E_1^2,E_2^2,E_3^2).
\label{e50}
\end{equation}
Correspondence between various components and $f^{\mu \nu}$ is standard:
space-space plaquettes are magnetic,  space-time plaquettes are electric.
The energy and action densities are respectively
\begin{eqnarray}
\epsilon = \frac{1}{2} ( E^2 + B^2),
\nonumber\\
\gamma   = \frac{1}{2} ( E^2 - B^2).
\label{e60}
\end{eqnarray}
Since the magnetic contribution turns out to be negative, there is
a strong cancellation between the two terms in the energy, whereas they
are enhanced in the action.
\begin{figure}
\begin{picture}(430,400)
\end{picture}
\caption{Energy and action profiles: (a) energy density in the plane
containing $q \bar{q}$;
(b) energy density in the plane midway between $q$ and $\bar{q}$;
(c) and (d) similarly for action density.}
\label{f1}
\end{figure}

Figure 1 gives the flux profiles.  The cancellation which suppresses
the energy density in the flux tube is evident.  However notice that
the self energy of the quarks is not similarly suppressed.  This
follows because the self energy is primarily electric.
We fitted the energy and action density in the plane at the midpoint
between $q$ and $\bar{q}$ using the function
\begin{eqnarray}
f(r_{\perp}) = a \exp\Bigl({-\sqrt{b^2 + (r_{\perp}/c)^2}}\Bigr).
\label{e70}
\end{eqnarray}
The peak value and the width at half maximum were very well determined
using a $\chi^2$ fit for each of 70 cases of different loop
sizes and values of $\beta$.
For the third parameter we chose the decay length of
the tail of this function and found it less well determined
but with a value typically close to the width at half maximum.
The details of the analysis will be given in a forthcoming paper\cite{hsw}.
Here we just give the results of the extrapolation to infinite
Wilson loop time extent in Figs. 2 and 3.
\begin{figure}
\begin{picture}(430,440)
\end{picture}
\caption{(a) Peak value of energy and action density; solid squares:
$\beta=2.5$;
triangles: $\beta=2.4$; open squares: $\beta=2.3$.  The two curves are
$1/R$ and $1/R^4$ arbitrarily normalized; (b) blowup of (a).}
\label{f2}
\end{figure}
\begin{figure}
\begin{picture}(430,220)
\end{picture}
\caption{Width at half maximum for energy and action density.}
\label{f3}
\end{figure}

The basic issue is whether the peak value of the energy density
stabilizes to a constant or goes to zero with quark separation.
This is not easy to settle as can be seen in Fig. 2. Roughly speaking
we know from the linearly rising potential that the string tension
$\sim$ (width)$^2$ $\times$ (peak value) should be constant.
Both Figs. 2 and 3 show that we are marginally asymptotic in
quark separation, $R$.
The two curves are $\sim 1/R^4$ and $\sim 1/R$.  A Coulomb field
would fall like the former but since the string tension is constant
the asymptotic width would have to grow like $R^2$ which clearly it does not.
Therefore we can rule out a Coulomb field as expected.
Interestingly for small separations,
the eyeball fit to  $\sim 1/R^4$ is quite good
which may be due to a Coulomb like behavior at small distances.
(The above argument that the width must grow like $R^2$ does not apply
because there is no string for small $R$.)
The dielectric model\cite{adler} predicts  the peak density $\sim 1/R$.
This function (arbitrarily normalized) does not
seem to fit the data very well.
However such a behavior would imply the width $\sim \sqrt{R}$
which is certainly possible in our data.  We can say that
the peak energy density and width are consistent with
a constant value for large quark separation but we can not rule out
a slow variation.  The issue can be tightened by making use of
the Michael sum rules\cite{michael1} as we mention in the next section.
\begin{figure}
\begin{picture}(430,220)
\end{picture}
\caption{Peak density for $R \times T$ Wilson loop sizes $R = 3-6$,
$T = 3-6$. For fixed $T$ the points decrease monotonically with $R$.
Triangles: energy density for
$R \times T$ loop; circles: energy density for
$T \times R$ loop; squares: $\|$ components of $E^2$ and $B^2$ only.}
\label{f4}
\end{figure}

Figure 4 illustrates a general feature of our data.
The cluster of three points
for each $R$ and $T$ correspond to the three quantities:
\begin{eqnarray}
\epsilon &=&
 \frac{1}{2} (E_{\|}^2 + B_{\|}^2)
         + \frac{1}{2} (E_{\perp}^2 + B_{\perp}^2),
\nonumber\\
\epsilon_{\|} &=& \frac{1}{2} ( E_{\|}^2 + B_{\|}^2),
\nonumber\\
\epsilon (T \leftrightarrow R)&=& \frac{1}{2} ( E_{\|}^2 + B_{\|}^2)
         - \frac{1}{2} (E_{\perp}^2 + B_{\perp}^2).
\label{e80}
\end{eqnarray}
If one turns the
Wilson loop on its side, the $\|$ components are unchanged but the
$\perp$ components of the electric
and magnetic fields are reversed:
$E_{\perp}^2 \leftrightarrow -B_{\perp}^2$. Hence there is a sign change
in the third expression.
The central points of the cluster are the $\|$ components only.
{\it The clustering of the points implies that the $\perp$ components
of the electric and magnetic contributions to energy density
are approximately equal but of opposite sign and cancel.}
The width of the peak is even less sensitive to the transverse components
giving essentially the same value for all three points.
%
\vglue 0.4cm
{\it \noindent 2.3. Sum Rules}
\vglue 0.1cm
A consistency check on the flux distributions can be obtained
by using the Michael sum rules\cite{michael1} for energy and action.
\begin{eqnarray}
 \frac{1}{2} \sum_{\vec{x}} ( E(\vec{x})^2 + B(\vec{x})^2 )&=& E_{0}(R) ,
\nonumber\\
\frac{1}{2}
\sum_{\vec{x}} ( E(\vec{x})^2 - B(\vec{x})^2 ) &=&  -\beta
\frac{\dot{a}}{a} [ E_{0}(R) - \frac{c(\beta)}{a}] -  \beta
\frac{\dot{c}(\beta)}{a}.
\label{e90}
\end{eqnarray}
Here $E_{0}(R)$ is given by Eqn.(\ref{e10}),
and ($ \cdot \equiv \frac{d}{d\beta}$).  In ref.\cite{hw} we have shown that
our data are essentially consistent with these sum rules.  The one difficulty
is the fact that the self energy,  $c(\beta)/a(\beta)$,
determined from the potential differs from the self energy determined
from the the action sum rule.  This may be due to an ambiguity in the
definition of self energy or possibly due to our classical
expressions for energy and action which ignores quantum corrections.
By taking a derivative of these expressions with respect to the
quark separation, R, this difficulty is avoided.
This gives the relation
\begin{equation}
\sigma_A = -\beta\frac{\dot{a}}{a} \sigma;
\;\;\;\;
\sigma_A \equiv
\frac{1}{2}\sum_{\vec{x}_{\perp}} ( E(\vec{x})^2 - B(\vec{x})^2 ); \\
\;\;\;\;
\sigma \equiv
\frac{1}{2}\sum_{\vec{x}_{\perp}} ( E(\vec{x})^2 + B(\vec{x})^2 ).
\label{e100}
\end{equation}
The sums are now over the plane midway between the $q \bar{q}$ pair.

Using the sum rules we find that the
$\beta$ function $(-\beta \dot{a}/a) \approx  10. \pm 2.$ compared
to the current estimates $ 7.\pm 1. $.
The asymptotic value is $-51/121 + 3\pi^2\beta/11
= 6.0 (\beta=2.4)$.  There is ample evidence from other measurements
that although scaling works well, asymptotic scaling is violated\cite{michael2}
and hence we do not expect to get the asymptotic value.

An alternative approach is to assume the sum rules are correct and use
them to infer information about the energy density from the action
density which is far easier to measure  since relative errors
are down by an order of magnitude.  As is clear from the sum rule,
the action does not scale yet the variation over these values of $\beta$
is very small.  An examination of Fig. 2(a) shows that the action
for each $\beta$ seems to stabilize to a constant for increasing
distance for the peak density and for the width.
This is quite striking for $\beta = $ 2.3 and 2.4.  For $\beta = $ 2.5
$R$ appears to be too small to draw a conclusion.  These data do not
suggest that the peak value is tending to zero at all.
We would like to use the sum rules to predict the behavior of the
energy density.  A constant peak energy density follows only
if the widths of the energy and action peaks have the same behavior.
Figure 3 shows that in fact they do.
{\it From this and using the sum rules we conclude that the energy density
stabilizes to a constant value also.}
This conclusion is an argument against the dielectric model\cite{adler}.
However we have little to say about logarithmic behavior of the
flux tube width as predicted by L\"{u}scher\cite{luscher}.  For
more details see Ref.\cite{hsw}.
%
\vglue 0.6cm
{\bf\noindent 3. Mechanism for confinement in U(1)}\footnote{Singh, Haymaker
and Browne}
\vglue 0.2cm
{\it\noindent 3.1. Dual Superconductor Model of Confinement}

\vglue 0.1cm
\baselineskip=14pt

We now turn to the mechanism for flux tube formation
and present direct evidence
that supercurrents of magnetic monopoles produce a dual Abrikosov
vortex\cite{monopoles}.
U(1) lattice gauge theory in 4 dimensions has both a confined phase at
large charge and a weak coupling deconfined phase corresponding to
continuum electrodynamics with a Coulomb interaction between static
charges.  Therefore confinement or its absence can be studied using U(1)
lattice gauge theory as a prototype, before tackling the more
complicated non-Abelian theories that actually describe quarks.  Much
evidence for the dual superconductor hypothesis has accumulated from
studies\cite{Polyakov,Banks,DeGrand} of
lattice gauge theory.  Polyakov\cite{Polyakov} and Banks, Myerson and
Kogut\cite{Banks} showed that U(1) lattice gauge theory in the
presence of a quark-antiquark pair could be approximately transformed
into a model describing magnetic current loops (the monopoles)
interacting with the electric current generated by the $q\bar{q}$
pair.  DeGrand and Toussaint\cite{DeGrand} demonstrated via a numerical
simulation that the vacuum of U(1) lattice gauge theory was populated
by monopole currents, copious in the confined phase and rare in the
deconfined phase.  This behavior has also been seen in non-Abelian
models after gauge fixing\cite{schierholz}.  Many studies of non-Abelian
models using Dirac monopoles\cite{schierholz,Suzuki} or other
topological excitations\cite{SmitIvanenko} support the dual
superconductor mechanism, although other studies\cite{NODSM} dissent.

So far, studies of confinement have examined ``bulk'' properties such
as the monopole density\cite{DeGrand,schierholz}, and the
behavior of the static quark potential\cite{Suzuki}.
In a recent paper\cite{shb} we presented
the first direct evidence that the flux tube is a dual Abrikosov vortex.
We further show that there
are exact U(1) lattice gauge theory analogues of two key
relations that lead to the Meissner effect in a
superconductor; the London equation and the fluxoid quantization
condition.

\vglue 0.4cm
{\it \noindent 3.2. Electric Field Profiles}

\vglue 0.1cm

Our simulations were done on a Euclidean spacetime lattice of volume
$9^3\times 10$.
The static charges are represented by a Wilson loop as in the previous
section. We take a $3\times3$ loop in the $z-t$ plane and measure the
fields in the $x-y$ plane at the midpoint between the charges.
Because of the geometrical symmetry of
the measurements only the $z$-components of $\langle\vec{{\cal
E}}\rangle$ and $\langle\vec{\nabla}\times \vec{J}_M\rangle$ are nonzero.  If
the Wilson loop is removed, even the $z$-components average to zero, so
the response is clearly induced by the presence of the static charges.
Only the imaginary part of the Wilson loop contributes to the averages
of these two quantities.

The plaquette measures flux passing through a unit
square on the lattice\footnote{The flux here means $E \times$
(area). We
use the same term for $E^2$ or $B^2$ since it has become an accepted usage.}
\begin{equation}
\exp[iea^2F_{\mu\nu}(\vec{r})] =
\exp[i\theta_{\mu\nu}(\vec{r})] \equiv
U_\mu(\vec{r})U_\nu(\vec{r}+\mu)U_\mu^{\dag}(\vec{r}+\nu)U_\nu^{\dag}(\vec{r}).
\label{110}
\end{equation}
The electric flux in lattice variables is
\begin{equation}
{\cal E}_{\mu}(\vec{r}) = {\rm Im } \exp[i\theta_{\mu4}(\vec{r})].
\label{111}
\end{equation}
\begin{figure}
\begin{picture}(430,250)
\end{picture}
\caption{Surface plot of the electric flux through the $xy$ plane
midway between the $q\bar{q}$ pair when the system is in (a) the
deconfined phase ($\beta=1.1$) and (b) the confined phase
($\beta=0.95$).  The line joining the pair is located at (0,0).}
\label{f5}
\end{figure}

Figure 5(a) shows the electric flux distribution for
$\beta=1.1$ where the vacuum is in the deconfined phase.  The broad
flux distribution seen is identical to the dipole field produced by
placing two classical charges at the quark positions, except that the
classical value of the flux on the $q\bar{q}$ axis is a factor of two
smaller.  We measure the total electric flux from one quark to the
other, including not only the flux through the plane between the
charges ($0.8504\pm0.0045$) but also the flux ($0.0951\pm0.0028$) that
flows through the lattice boundary because of the periodic boundary
conditions.  This yields a total flux of $(0.9453\pm0.0053$), close to
the theoretical value $\Phi_e = e/\sqrt{\hbar c} = 1/\sqrt{\beta} =
0.9534$.

\begin{figure}
\begin{picture}(430,500)
\end{picture}
\caption{
Behavior of (a) the electric flux, (b) the curl of the monopole current
and (c) their sum, the fluxoid, in the confined phase ($\beta=0.95$) as
a function of the perpendicular distance $r_{\perp}$ from the $q\bar{q}$ axis.
The solid line in (a) gives the flux for a pure Coulomb field .}
\label{f6}
\end{figure}

Figure 5(b) and 6(a) shows the electric flux in the confined phase
($\beta=0.95$).  In this case the flux is confined almost
entirely within one lattice spacing of the axis and essentially no flux
passes the long way around through the lattice boundary.  The
net flux is again equal to $1/\sqrt{\beta}$ within statistical
error.  This behavior is exactly what one would expect from the
superconducting analogy, where the flux has been ``squeezed'' into a
narrow tube.
\vglue 0.4cm
{\it \noindent 3.3. Magnetic Monopole Supercurrents}
\vglue 0.1cm

The monopole currents are found by a prescription devised by
DeGrand and Toussaint\cite{DeGrand}, which employs a lattice version of
Gauss' Law to locate the Dirac string attached to the monopole.  The
net flux into each plaquette face is given by
$(\theta_{\mu\nu}(\vec{r}) \bmod 2\pi)$.  If the sum of the fluxes into
the faces of a 3-volume at fixed time is nonzero, a monopole is located
in the box. A non-zero net flux can occur only if a multiple of $2\pi$
arises from the mod operation.  Therefore the net flux is
\begin{equation}
\sum_{6 faces}a^2 F_{\mu\nu} = n (2 \pi \hbar c/ e) = n e_M
\label{120}
\end{equation}
The net flux into the box at fixed time thus yields the
monopole ``charge'' density, the time component of the monopole
4-current $J_M$. The spatial components are found similarly.  The
monopole currents form closed loops due to the conservation of
magnetic charge. Finally the curl is calculated by the line integral of
the current around a dual plaquette.

We show in Fig. 6(a) $\langle\vec{\cal E}\rangle$ and in
Fig. 6(b) $-\langle\vec{\nabla}\times\vec{J}_M\rangle$ in the
confined phase as a function of the distance from the $q\bar{q}$ axis.
The data show that the spatial variation of the flux and the curl of
the current are very similar, except for the point on the axis which
will be discussed below.

\vglue 0.4cm
{\it \noindent 3.4. London Equations,
Fluxoid Quantization and the Abrikosov Vortex}

\vglue 0.1cm
In order to interpret this result we would like to review the ordinary
London theory. In the next section we interchange electric
and magnetic quantities to get the dual results.

A concise statement of the London theory is contained
in the relation\footnote{We use Heaviside-Lorentz units to be consistent with
lattice gauge theory.}
\begin{equation}
\vec{A} + \frac{\lambda^2}{c} \vec{J}= 0; \;\;\;\;
(\vec{\nabla} \cdot \vec{A} = 0).
\label{121}
\end{equation}
If the charge density is zero, then in this gauge the electric field
is given by $-\dot{A}/c$ and therefore
$\vec{E} = \lambda^2 \dot{\vec{J}}/c$.
This describes a perfect conductor and is just Newton's law for
free carriers, $e \vec{E} = m \dot{\vec{v}}$.  By taking the curl
of Eqn.(\ref{121}) we obtain the condition for a perfect diamagnet.
\begin{equation}
\vec{\nabla} \times \vec{J}  = - \frac{c}{\lambda^2} \vec{B}.
\label{140}
\end{equation}
This relation together with Ampere's law
$\vec{\nabla} \times \vec{B} = \vec{J}/c$ gives
$\nabla^2 \vec{B} = \vec{B}/\lambda^2$ which implies that the magnetic field
falls off in the interior of the superconductor with a skin depth $\lambda$.

Finally the fluxoid is given by the integral
\begin{equation}
\int_{S} (\vec{B}+ \frac{\lambda^2}{c}\vec{\nabla} \times \vec{J})\cdot \hat{n}
da =
\int_{C} (\vec{A}+ \frac{\lambda^2}{c}\vec{J})\cdot d\vec{l} =
n \frac{\hbar c}{2e} = n \Phi_{m}
\label{150}
\end{equation}
If the curve $C$ is in a simply connected region of a superconductor, then
$n = 0$.  However if the curve encircles a hole in the material then
n need not be zero but must be an integer.  In an extreme type II
Abrikosov vortex, a very
small core is comprised of normal material.  A single unit of magnetic
flux of radius $\sim \lambda$ passes through the vortex.  The fluxoid
density $\vec{B} + \frac{\lambda^2}{c}\vec{\nabla} \times \vec{J}$ is zero
everywhere except in the region of the normal material.  In the limit
in which the core is a delta function we obtain:
\begin{equation}
\vec{B} + \frac{\lambda^2}{c}\vec{\nabla} \times \vec{J}=
\Phi_{m} \delta^2(x_{\perp})\hat{n}_z.
\label{160}
\end{equation}
Further if we use Ampere's law we can get an analytic expression for $B_z$.
\begin{equation}
B_z(r_{\perp}) - \lambda^2\nabla^2 B_z= \Phi_{m} \delta^2(x_{\perp});
\;\;\;\; B_z = \frac{\Phi_{m}}{2 \pi \lambda^2} K_{0}(r_{\perp}/\lambda).
\label{170}
\end{equation}
%

\vglue 0.4cm
{\it \noindent 3.5. Dual Superconductor }

\vglue 0.1cm
We interpret our results using the following relations which are
the dual of the corrsponding relations in the previous section:
\begin{equation}
\vec{E} - \frac{\lambda^2}{c}\nabla \times \vec{J_m}
 = \Phi_{e} \delta^2(x_{\perp})\hat{n}_z.
\label{190}
\end{equation}
where the unit of electric flux  $\Phi_{e} = 1/\sqrt{\beta}$.
Figure 5(c) shows the result of fitting our data to this relation.
The London penetration depth, $\lambda$, is the only free parameter.
We are able
to determine a value of $\lambda$ which gives zero away from the axis and a
delta function with the correct coefficient on axis.  Further
we can check that the electric flux profile is given by
\begin{equation}
E_z = \frac{\Phi_{e}}{2 \pi \lambda^2} K_{0}(r/\lambda)
\label{200}
\end{equation}
This function has no free parameters.
This curve is also shown in Fig. 6(a)  showing excellent agreement.
We find a
value of $\lambda/a = 0.482\pm0.008$, which is consistent with the range
of penetration of the electric flux in Fig. 6(a) and the
thickness of the current sheet in Fig. 6(b).  We expect
that, as in a superconductor, the transition to the deconfined phase
will be signalled by a divergence of the London penetration depth.  We
have therefore measured $\lambda$ further from the deconfinement
transition at $\beta=0.90$, and find a smaller penetration depth of
$\lambda/a = 0.32\pm0.02$.  In the deconfined phase, $\beta=1.1$
we find an almost
insignificant value of $\langle\vec{\nabla}\times \vec{J}_M\rangle$ and
fitted values of $\lambda$ were larger than our lattice size.

In summary, one can find a value of the London penetration depth,
$\lambda$, that
satisfies Eqn.(\ref{190}) off axis.  One then finds that for the point
on axis, the same value of $\lambda$ gives one quantum of electric
flux as predicted by Eqn.(\ref{190}).  Finally the same value of
$\lambda$ gives a good fit to the profile using Eqn.(\ref{200}).
Hence considerable detail of the dual Abrikosov is verified. It is
perhaps surprising that a nonlinear, strongly interacting, model such
as U(1) lattice gauge theory could be described by such a simple model
as the linear London equations but our results indicate that the
operators
$\langle\vec{{\cal
E}}\rangle$ and $\langle\vec{\nabla}\times \vec{J}_M\rangle$
when measured in the presence of source of external flux like a Wilson
loop, give an unambiguous indication of the confinement of electric
flux by a monopole current distribution.  The simulation yields a large
signal even with modesnt amounts of computer time on a Sun workstation.
Although the Meissner effect itself requires only that
Eq.~(\ref{200}) hold off axis, our data also support the more restrictive
fluxoid quantization relation on axis.
This additional relation reflects the single-valued nature of the
order parameter in a Ginzburg-Landau description of the monopole
condensate.    Because the
monopoles appear pointlike in our simulations, lattice gauge theory
looks like an extreme type-II superconductor.

\vglue 0.6cm
{\bf \noindent 4. Acknowledgements \hfil}

\vglue 0.4cm
Two of us (R.W.H. and V.S.) would like to thank A. Kotanski,
Y. Peng, G. Schierholz and T. Suzuki for many
fruitful discussions on this problem.
V.S. thanks A. Kronfeld and
R. Wensley for useful conversations.  R.W.H and V.S. are
supported by the DOE under grant DE-FG05-91ER40617 and D.A.B. is
supported in part by the National Science Foundation under Grant No.
NSF-DMR-9020310 J.W. is supported in part by the Polish Government
under Grants Nos. CPBP 01.01 and CPBP 01.09.

\vglue 0.6cm
{\bf\noindent 5. References \hfil}

\vglue 0.4cm

\vglue 0.2cm

\newpage

a

\bigskip

b

\bigskip

c

\bigskip

d

\bigskip

a

\bigskip

b

\bigskip
\bigskip
\bigskip
\bigskip

LSU HE No. 130-1992

\end{document}